\documentclass[twocolumn,aps,prl,10pt]{revtex4-1}

\usepackage{amsmath}
\usepackage{amssymb}
\usepackage{graphicx}
\usepackage{hyperref}
\usepackage{dsfont}
\usepackage{txfonts}

\hypersetup{
    colorlinks,
    linkcolor={blue},
    citecolor={blue},
    urlcolor={blue}
}

\allowdisplaybreaks

\graphicspath{{./}{./figs/}{../figs/}{../Figs_trial/}{../Yukawa/}}

\newcommand{\rme}{\mathrm{e}}

\newcommand{\llangle}{\langle\!\langle}
\newcommand{\rrangle}{\rangle\!\rangle}

\renewcommand{\vec}[1]{\boldsymbol{#1}}

\usepackage[dvipsnames]{xcolor}
\usepackage[normalem]{ulem}

\begin{document}

\title{%
Soluble Fermionic Quantum Critical Point in Two Dimensions
}

\author{Shouryya Ray}
\author{Matthias Vojta}
\author{Lukas Janssen}
\affiliation{Institut f\"ur Theoretische Physik, TU Dresden, 01062 Dresden, Germany}
\affiliation{W\"urzburg-Dresden Cluster of Excellence ct.qmat, TU Dresden, 01062 Dresden, Germany}

\begin{abstract}
We study a model for a quantum critical point in two spatial dimensions between a semimetallic phase, characterized by a stable quadratic Fermi node, and an ordered phase, in which the spectrum develops a band gap.
The quantum critical behavior can be computed exactly, and we explicitly derive the scaling laws of various observables.
While the order-parameter correlation function at criticality satisfies the usual power law with anomalous exponent $\eta_\phi = 2$,  the correlation length and the expectation value of the order parameter exhibit essential singularities upon approaching the quantum critical point from the insulating side, akin to the Berezinskii-Kosterlitz-Thouless transition. The susceptibility, on the other hand, has a power-law divergence with non-mean-field exponent $\gamma = 2$. %
On the semimetallic side, the correlation length remains infinite, leading to an emergent scale invariance throughout this phase.
%
%
\end{abstract}

\date{\today}

\maketitle


Quantum critical matter exhibits exotic behavior that substantially differs from the situation in stable phases~\cite{sachdev2011}.
In stable metallic phases, for instance, the usual effective excitations are Fermi quasiparticles; insulating magnets are, as long as not highly frustrated, describable in terms of weakly-interacting magnons. In the vicinity of quantum critical points (QCPs), however, novel types of excitations emerge: They may be quasiparticle excitations with fractionalized quantum numbers~\cite{senthil2004}, or they may not admit a quasiparticle description at all~\cite{loehneysen2007}. Both cases lead to fascinating phenomena that are only poorly understood to date.
This is particularly true for low-dimensional Fermi systems, which pose a serious challenge to theory. The reason for the complications is the fact that integrating out the fermion modes, as is done in the standard Hertz-Millis theory~\cite{hertz1976, millis1993}, may lead to nonanalytic terms in the effective action for the critical order-parameter modes~\cite{belitz2005}. This invalidates the Gaussian approximation and modifies the critical behavior, or might even render the transition discontinuous~\cite{kirkpatrick2003, rech2006}.
A proper description of a QCP in a two-dimensional metallic system generically requires an infinite set of $(2+1)$-dimensional local field theories, labeled by points on the Fermi surface; in order to gain analytical control over the problem, the Fermi surface needs to be restricted to a finite number of ``patches''~\cite{lee2009, metlitski2010a, metlitski2010b, holder2015}.

The problem becomes slightly less difficult in situations when the Fermi surface shrinks to isolated points in the Brillouin zone. Such a situation arises in the $d$-wave superconducting state on the square lattice, in which case the Fermi surface comprises four Dirac nodes. The onset of order out of this state is then described by a $(2+1)$-dimensional relativistic Gross-Neveu-Yukawa field theory~\cite{vojta2000}. Similarly, QCPs in models for monolayer graphene are described by variants of the Gross-Neveu-Yukawa theory~\cite{herbut2006, herbut2009}. In Bernal-stacked bilayer graphene, which in the simplest tight-binding description hosts a quadratic band touching point at the Fermi level, interactions generically induce Dirac nodes in the spectrum~\cite{pujari2016, leaw2019}; this leads to emergent relativistic symmetry and a Gross-Neveu QCP in the phase diagram~\cite{ray2018}. Interacting Dirac fermions on the surface of a three-dimensional topological insulator can realize a variant of the Gross-Neveu quantum universality class that is characterized by emergent supersymmetry, relating the fermionic degrees of freedom with the composite bosonic order-parameter field~\cite{lee2007, grover2014, jian2015, jian2017, zerf2016}.

Although the Gross-Neveu-Yukawa problem has received significant attention recently, no definite quantitative consensus on the critical behavior of many of these universality classes has been reached: On the analytical side, divergences in the series expansions impede a simple extrapolation to the physically relevant cases~\cite{zerf2017, ihrig2018}; on the numerical side, the available lattice sizes in quantum Monte Carlo simulations of models with gapless fermions are significantly smaller than those reachable in purely bosonic systems~\cite{assaad2013, li2015, otsuka2016, buividovich2018, lang2019}.
To the best of our knowledge, an exactly soluble QCP in a clean two-dimensional many-body Fermi system is not yet known~\cite{endnote1}.

In this Letter, we present a welcome counterexample to these notorious difficulties.  We propose a simple model of interacting electrons on the kagome lattice that realizes a stable QCP that is \emph{not} of the Gross-Neveu type. At criticality, the system flows to a Gaussian renormalization group (RG) fixed point of the corresponding continuum field theory. This allows a full solution of the quantum critical behavior, which, despite being Gaussian, turns out to be nontrivial, with some observables showing the usual power-law behavior, while others exhibit essential singularities.
For instance, the correlation function of the order parameter has a power-law form with a finite anomalous dimension $\eta_\phi = 2$ and dynamical exponent $z = 2$.
Similarly, when the QCP is approached from the insulating side, the magnetic susceptibility scales as a power law with susceptibility exponent $\gamma = 2$.
By contrast, the correlation length diverges in the form of an essential singularity, corresponding formally to a divergent correlation-length exponent $\nu = \infty$.
Furthermore, on the semimetallic side of the transition, the correlation length remains infinite, and we demonstrate that this phase is characterized by emergent scale invariance with power-law correlation functions---a critical phase.

Interacting Fermi systems with quadratic band touching have been intensely studied before. In three dimensions, weak short-range interactions are power-counting irrelevant, but the long-range Coulomb interaction may induce a quantum critical gapless phase with nontrivial power-law exponents \cite{abrikosov1974, moon2013}, or an insulating state with topologically protected gapless edge modes \cite{herbut2014, janssen2016a, janssen2017}. In two dimensions, the long-range interaction is screened and most previous work focused on repulsive short-range interactions, for which the system is believed to be unstable towards the formation of a quantum anomalous Hall (QAH) state \cite{sun2009, uebelacker2011, murray2014, zhu2016, zeng2018, ren2018}; a recent numerical study also considers attractive interactions, suggesting the presence of a stable semimetallic phase \cite{sur2018}. Here, we argue that this gapless phase is quantum critical (albeit with integer power-law exponents) and that the interaction-induced semimetal-insulator transition is exactly solvable.

\paragraph{Lattice model.}

\begin{figure}
  \includegraphics[width=\linewidth]{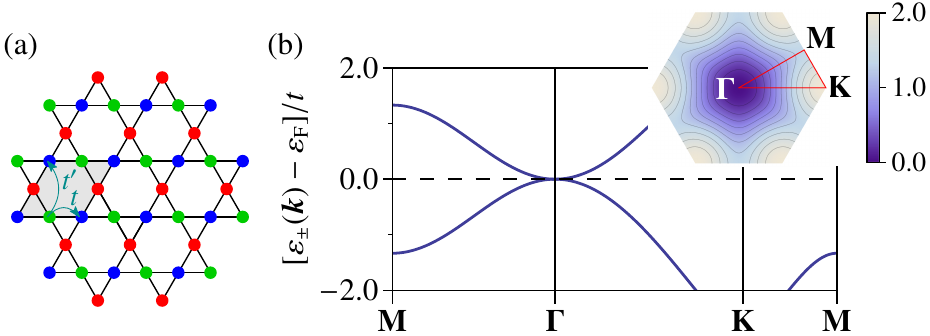}
  \caption{(a)~Kagome lattice. The unit cell (shaded rhombus) consists of three sites (red, green, and blue dots). (b)~Spectrum of the noninteracting Hamiltonian $H_0$ (for $t' = -t/3$) along high-symmetry lines, displaying the quadratic band touching point at the center of the Brillouin zone. The dashed line denotes the Fermi level for $2/3$-filling. Inset: Conduction-band dispersion in the first Brillouin zone (color plot) and path used in the main panel (red line).}
  \label{fig:kagome}
\end{figure}

\begin{figure}
\includegraphics[width=\linewidth]{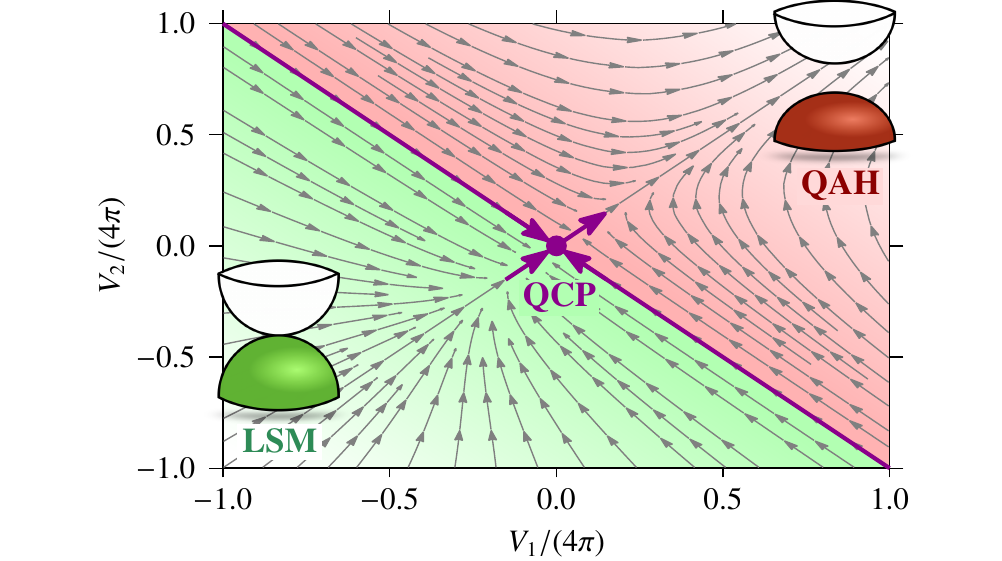}
\caption{%
RG flow of interacting spinless fermions on the kagome lattice at $2/3$-filling in the space of nearest-neighbor interaction $V_1$ and next-nearest-neighbor interaction $V_2$, in units of $t=2$.
For dominating repulsive interactions, $V_1 > - V_2$, the system becomes a quantum anomalous Hall (QAH) insulator, while it realizes a Luttinger semimetal (LSM) with emergent scale invariance for dominating attractive interactions, $V_1 < -V_2$.
The critical behavior of the continuous semimetal-insulator transition is governed by the Gaussian quantum critical fixed point (QCP).
}
\label{fig:V1V2phasediag}
\end{figure}

For concreteness, consider a model of interacting spinless fermions on the kagome lattice, defined by the Hamiltonian $H = H_0 + H_\mathrm{int}$. The noninteracting part reads
\begin{align}
 H_{0} & = - t\sum_{\langle i j \rangle} c_i^\dagger c_j - t' \sum_{\llangle i j \rrangle} c_i^\dagger c_j + \textnormal{H.c.},
\end{align}
where $t$ ($t'$) is the hopping amplitude for nearest (next-nearest) neighbors $\langle ij \rangle$ ($\llangle ij \rrangle$) on the kagome lattice, with $c_i^\dagger$ and $c_i$ denoting fermion creation and annihilation operators at site~$i$; Fig.~\ref{fig:kagome}(a).
We assume a filling fraction of $2/3$ for which the Fermi surface shrinks to an isolated point at $\vec{q} = 0$, with a quadratic quasiparticle dispersion $\varepsilon_\pm({\vec{q}}) - \varepsilon_{\text{F}} = \tfrac12[-(t+3t')\pm(t-3t')] \vec{q}^2$ and Fermi energy $\varepsilon_{\text{F}} = 2t(1 + t'/t)$; Fig.~\ref{fig:kagome}(b).
For simplicity, we choose in the following $t'/t = -1/3$, yielding a particle-hole-symmetric spectrum $\varepsilon_\pm({\vec{q}}) - \varepsilon_{\text{F}} = \pm t \vec{q}^2$. Particle-hole symmetry in fact turns out to be emergent at criticality~\cite{suppl}. We assume density-density interactions on nearest- and next-nearest neighbor bonds,
\begin{align}
 H_\mathrm{int} &= V_1 \sum_{\langle i j \rangle} c_i^\dagger c_i c_j^\dagger c_j  + V_2 \sum_{\llangle i j \rrangle} c_i^\dagger c_i c_j^\dagger c_j,
\end{align}
where $V_{1,2} > 0$ ($V_{1,2} < 0$) corresponds to repulsive (attractive) interactions.
Fig.~\ref{fig:V1V2phasediag} shows the low-temperature phase diagram of the model as function of $V_1$ and $V_2$, as computed below.

\paragraph{Low-energy field theory.}

The low-energy physics is governed by small momenta near the quadratic Fermi node, described by the effective Euclidian action $S = \int \mathrm{d}\tau \mathrm{d}^2 \vec{x} \mathcal{L}$, with
\begin{align}
  \mathcal{L} & = \psi^\dagger\left[\partial_\tau - \left(\partial_x^2 - \partial_y^2\right)\sigma_3 - 2 \partial_x \partial_y \sigma_1 \right]\psi - \frac{g}{2}(\psi^\dagger \sigma_2 \psi)^2 + \dots
\label{eq:lowEtheorymaster}
\end{align}
In the above equation, the ellipsis denote higher-order terms of the form $\sim g' (\psi^\dagger \sigma_\alpha \psi)\partial_i \partial_j (\psi^\dagger \sigma_\beta \psi)$, $i,j \in \{x,y\}$, $\alpha, \beta \in \{0,1,2,3\}$, which are irrelevant in the RG sense,
$\sigma_0 \equiv \mathds 1_2$, and $\sigma_1, \sigma_2, \sigma_3$ are the standard Pauli matrices. The components of the fermion field
$\psi = (\psi_1, \psi_2)^\top$ are given by the annihilation operators $c_{i\mathrm A}$, $c_{i\mathrm B}$, and $c_{i\mathrm C}$ on the three triangular sublattices A, B, and C as %
$c_{i\mathrm A} \to \frac{1}{\sqrt{6}} \psi_1 - \frac{1}{\sqrt{2}} \psi_2$, %
$c_{i\mathrm B} \to -\sqrt{\frac{2}{3}} \psi_1$, %
and $c_{i\mathrm C} \to \frac{1}{\sqrt{6}} \psi_1 + \frac{1}{\sqrt{2}} \psi_2$.
The four-fermion coupling $g$ is determined by the nearest- and next-nearest neighbor interaction strengths as $g \simeq 2(V_1 + V_2)/t$, while the complementary combination $(V_1 - V_2)$ corresponds to a higher-order gradient term in the Lagrangian, $g' \simeq V_1 - V_2$; see Ref.~\cite{suppl} for details.
%
%
Note that $g$ is dimensionless and hence corresponds to a marginal operator, while $g'$ is power-counting irrelevant. We emphasize that Eq.~\eqref{eq:lowEtheorymaster} includes all possible relevant or marginal terms compatible with the symmetry of the lattice model.
%
For notational simplicity, we choose in the following units in which $t = 2$, such that $g \simeq V_1 + V_2$.

\paragraph{RG flow and phase diagram.}

Integrating out high-energy modes with momenta $q \in (\Lambda/b, \Lambda)$ and all frequencies $\omega \in (-\infty, \infty)$ causes the coupling constants to flow according to
\begin{align}
  \frac{d g}{d \ln b} & = \frac{g^2}{4\pi}, &
  \frac{d g'}{d \ln b} & = - 2 g',
\end{align}
valid for small $|g|, |g'| \ll 1$, in agreement with the previous result for $g'=0$~\cite{sun2009}. Upon identifying $g \simeq V_1+V_2$ and $g' \simeq V_1 - V_2$ at the cutoff scale, the RG flow leads to the phase diagram depicted in Fig.~\ref{fig:V1V2phasediag}. %
The coupling plane is divided into two phases separated by the line $(V_1/V_2)_\mathrm{c} = -1$. For $V_1 > - V_2$, which corresponds to dominating repulsive interactions, the couplings diverge at the RG scale $\Lambda_{\text{SSB}} = \Lambda \rme^{-4\pi/g^{(0)}}$, where $g^{(0)}$ denotes the initial value of the marginal coupling at the ultraviolet scale. The divergence signals an instability of the semimetal towards an ordered ground state. %
A simple mean-field analysis~\cite{sun2009} suggests that among the various possible orderings, the QAH state~\cite{haldane1988} with order parameter $\langle \psi^\dagger \sigma_2 \psi \rangle$ has the lowest energy. This result is consistent with numerical works \cite{zhu2016,zeng2018,sur2018,ren2018}. The ordered phase is characterized by spontaneous time-reversal symmetry breaking, topologically protected gapless edge modes and a full gap in the bulk spectrum.
For dominating attractive interactions with $V_1 < -V_2$, however, the couplings flow to the Gaussian fixed point at $V_1 = V_2 = 0$, demonstrating the existence of a stable semimetallic phase with quadratic quasiparticle dispersion. An equivalent semimetallic phase has been found recently on the checkerboard lattice~\cite{sur2018}.
As the semimetallic phase is described by the two-dimensional version~\cite{janssen2015} of the Luttinger Hamiltonian~\cite{luttinger1956}, we dub this phase ``Luttinger semimetal'' (LSM) in analogy to the three-dimensional case~\cite{moon2013, herbut2014, janssen2016a, janssen2017, boettcher2016, boettcher2017}.
On the phase transition line, the Gaussian fixed point is RG attractive. The semimetal-insulator transition is therefore continuous. Here, we are interested in its critical behavior, as well as the properties of the LSM phase, both of which turn out to be nontrivial.

\paragraph{Critical behavior.}

\begin{figure}[t]
 \centering
 \includegraphics[width=\linewidth]{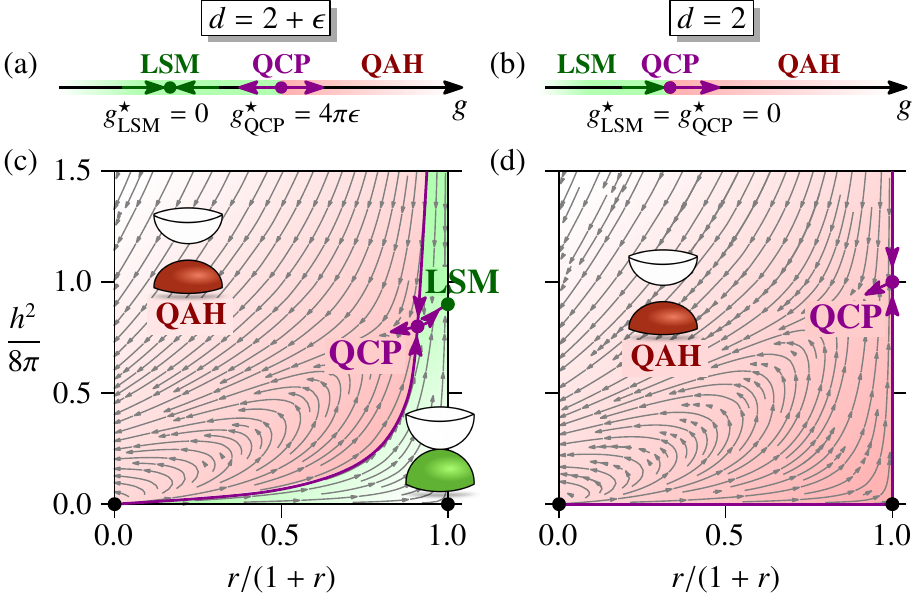}
 \caption{RG flow in the four-fermion model (a,b) and the Luttinger-Yukawa model for $c = 1/4$ (c,d) in $d=2+\epsilon$ (a,c) and $d=2$ (b,d) dimensions. The two fixed points at  $g^\star_\text{LSM}=0$ and $g^\star_\text{QCP}=4\pi\epsilon$ in the four-fermion model correspond to the nontrivial fixed points denoted as LSM and QCP, respectively, in the Luttinger-Yukawa model. For $\epsilon \to 0$, the two fixed points approach each other in both theories.}
\label{fig:h2rflow}
\end{figure}

For simplicity, we restrict ourselves in what follows to the case $V_1 = V_2$, for which we are left with a single parameter $g = V_1 + V_2$.
It is advantageous to first consider an extended theory in general spatial dimension $d = 2+\epsilon$ with $\epsilon \ll 1$ and postpone the limit $\epsilon \to 0$ until the end.
This technical trick has two advantages:
Firstly, it reveals that the quantum critical behavior can be understood as originating from a collision of two fixed points:
The four-fermion flow equation $(d g)/(d \ln b) = - \epsilon g + g^2/(4\pi)$ exhibits besides the infrared stable Gaussian fixed point $g^\star_\text{LSM}=0$ an ultraviolet stable interacting fixed point at $g^\star_\text{QCP} = 4\pi \epsilon$, see Fig.~\ref{fig:h2rflow}(a,b).
While the Gaussian fixed point governs the low-energy behavior of the LSM phase, the interacting fixed point controls the universal behavior upon approaching the critical point from the insulating side.
For instance, the corresponding correlation-length exponent can be readily obtained at this critical point as $\nu = 1/\epsilon+\mathcal O(\epsilon^0)$.
The interacting fixed point approaches the Gaussian fixed point for small~$\epsilon$ and collides with it in the limit $\epsilon \to 0$.
Such a collision of fixed points occurs generically in gauge theories, in which case the fixed points typically disappear into the complex plane after the collision \cite{herbut2014, janssen2017, gies2006, kaplan2009, braun2014, janssen2016b, herbut2016b, halperin1974, nahum2015a, ihrig2019, subils2019}. A collision of fixed points can also occur in systems without gauge invariance, with and without fixed-point complexification \cite{gehring2015, herbut2016a, roscher2018, gracey2018, rychkov2018}.
In all cases, the fixed-point collision leads to essential singularities in various observables, which is what we also find below.
The second advantage of our dimensional generalization is that it allows us to explicitly demonstrate the emergent scale invariance of the LSM phase.
For finite $\epsilon$, we may restrict to $g\geq 0$. By means of a Hubbard-Stratonovich transformation, the four-fermion interaction $(\psi^\dagger \sigma_2 \psi)^2$ can then be rewritten in terms of a Yukawa interaction between the fermion bilinear $\psi^\dagger \sigma_2 \psi$ and the order-parameter field $\phi$, with the fermion-boson Lagrangian
\begin{align}
  \mathcal{L}' & =
  \psi^\dagger\left[\partial_\tau - \left(\partial_x^2 - \partial_y^2\right)\sigma_3 - 2 \partial_x \partial_y \sigma_1 \right]\psi
  \nonumber \\ & \quad
  + \frac12 \phi \left(r - c \partial_\tau^2 - \partial_x^2 - \partial_y^2\right) \phi
   - h \phi \psi^\dagger \sigma_2 \psi,
\end{align}
which we dub ``Luttinger-Yukawa'' model.
Here, $r$ is the tuning parameter of the transition (boson mass), $h$ represents the Yukawa coupling, and the parameter $c$ accounts for the different scaling between spatial and temporal coordinates~\cite{janssen2015}.
The Luttinger-Yukawa model is equivalent to the four-fermion theory~\eqref{eq:lowEtheorymaster} upon the identification $g \equiv h^2/r > 0$. In particular, the order parameter is then given by $\langle\phi\rangle = (h/r)\langle\psi^\dagger \sigma_2 \psi\rangle$.
The RG flow in the plane spanned by the dimensionless parameters $r/\Lambda^2 \mapsto r$ and $h^2/\Lambda^2 \mapsto h^2$ is depicted in Fig.~\ref{fig:h2rflow}(c,d), where we have set the dimensionless parameter $c \Lambda^2 \mapsto c$ to its infrared stable fixed-point value $c^\star=\frac{1}{4} + \mathcal O(\epsilon)$~\cite{suppl}.

For $\epsilon > 0$, Fig.~\ref{fig:h2rflow}(c), there are two nontrivial fixed points at finite $h^2$, denoted as QCP and LSM.
The former has a unique RG-relevant direction, and it hence governs the continuous semimetal-insulator transition.
This fixed point is dual to the ultraviolet stable fixed point at $g^\star_\text{QCP} = 4\pi \epsilon$ in the four-fermion theory, which can be seen as follows:
The separatrix connecting this critical fixed point with the trivial fixed point at the origin (purple curve) is characterized by the fixed ratio $(h^2/r)^\star_\text{QCP} = 4\pi\epsilon$ for all $r$, in agreement with our previous result for $g^\star_\text{QCP}$.
Furthermore, we find the correlation-length exponent at this critical point as $\nu = 1/\epsilon$, which is again consistent with the four-fermion result.
Since $h^2$ is finite at this fixed point, we obtain a finite boson anomalous dimension $\eta_\phi = 2 - 2\epsilon$.
By contrast, the fermion anomalous dimension vanishes, $\eta_\psi = O(\epsilon^3)$, and the dynamical critical exponent is $z = 2 + \mathcal O(\epsilon^2)$.
Note that the fixed point is characterized by a large boson mass $r \gg 1$, implying that higher loop orders in the perturbative expansion do not contribute for small~$\epsilon$, despite that $h^{2}$ is finite.

\begin{table}[t]
\caption{Scaling behaviors at the QCP slightly above and directly in two dimensions. Here, $\delta g \equiv g - g_\mathrm{c} \geq 0$ measures the distance from criticality in the ordered phase and $H$ denotes an external field that couples to the order parameter.}
\label{tab:critexps}
\begin{tabular*}{\linewidth}{@{\extracolsep{\fill} }c c c c c c c c c}
  \hline\hline
  $d$ & $\langle \phi(0,\vec{x})\phi(0,0) \rangle$  & $\langle \phi(\tau,0)\phi(0,0) \rangle$  & $\xi$ & $\chi$ & $\langle \phi \rangle_{\delta g}$ & $\langle \phi \rangle_{H}$\\
  \hline
  $2 + \epsilon$ & $|\vec x|^{-4 + 2\epsilon}$ & $|\tau|^{-2 + \epsilon}$ & $\delta g^{-1/\epsilon}$ & $\delta g^{-2+\mathcal O(\epsilon)}$ & $\delta g^{2/\epsilon}$ & $H^{1+\epsilon}$ \\
  $2$ & $|\vec x|^{-4}$ & $|\tau|^{-2}$ & ${\rme}^{4\pi/\delta g}$ & $\delta g^{-2}$ & ${\rme}^{-8\pi/\delta g}$ & $H$ \\ \hline\hline
\end{tabular*}
\end{table}

In the four-fermion model, the Gaussian fixed point at $g_\text{LSM}^\star = 0$ is infrared stable. It therefore does not map to any of the trivial fixed points in the Luttinger-Yukawa theory, but rather to the fully stable nontrivial fixed point denoted as LSM. This can also be seen by considering the limit $\epsilon \to 0$: Upon lowering $\epsilon$, the two nontrivial fixed points approach each other and eventually merge for $\epsilon = 0$, Fig.~\ref{fig:h2rflow}(d). Now, all interacting flow lines cross the $r=0$ axis at finite $h^2$, signaling the instability of the LSM state for all $h^2/r = g>0$.
While the $\epsilon \to 0$ limit of the anomalous dimensions and the dynamical exponent can be taken in a straightforward manner, the correlation-length exponent $\nu$, and consequently also the exponents $\alpha$ and $\beta$ as obtained from hyperscaling, diverge in this limit.
To elucidate the meaning of these divergent exponents, we compute the effective potential directly in $d=2$, yielding~\cite{suppl}
\begin{align}
  V_{\text{eff}}(\phi)
  & = \frac{h^2\phi^2}{2h^2/r}\left[1 + \frac{h^2/r}{8\pi}\left(\ln\frac{h^2\phi^2}{4}-1\right)
+ \left(\frac{h^2/r}{16 \pi}\ln\frac{h^2\phi^2}{4}\right)^2\right],
\end{align}
where we have rescaled $V_\text{eff}/\Lambda^4 \mapsto V_\text{eff}$, $r/\Lambda^2 \mapsto r$, $h/\Lambda \mapsto h$, and $\phi/\Lambda \mapsto \phi$, and assumed $r \gg 1$, as before. This yields the order parameter
%
$\langle \psi^\dagger \sigma_2 \psi \rangle \propto \langle \phi \rangle \propto \rme^{-8\pi/g}$, if $g>g_\mathrm{c}=0$,
%
where we have re-identified $h^2/r = g$. Note that this is now an essential singularity in $g$, akin to the classical Berezinskii-Kosterlitz-Thouless (BKT) critical point~\cite{herbut2007}. A similar singularity presents itself for the correlation length,
%
$\xi \propto \rme^{4\pi/g}$, if $g>g_\mathrm{c}$.
%
For the free energy density, we analogously obtain $V_\text{eff} (\langle\phi\rangle) \propto \rme^{-16\pi/g}$, which confirms the hyperscaling assumption $V_\text{eff} (\langle\phi\rangle) \propto \xi^{-(d+z)}$ with $d=2$ and $z=2$ near criticality.
Since $\eta_\phi = 2$, and in contrast to the classical BKT transition, the magnetic susceptibility has the usual power-law divergence $\chi \propto g^{-\gamma}$ with $\gamma = 2$ for $g>g_\mathrm{c}$. Note that this result is different from the mean-field calculation, in which case one obtains $\gamma_\text{MF} = 1$.
The scaling behaviors of different observables are summarized in Tab.~\ref{tab:critexps}.

\begin{figure}[t]
 \centering
 \includegraphics[width=\linewidth]{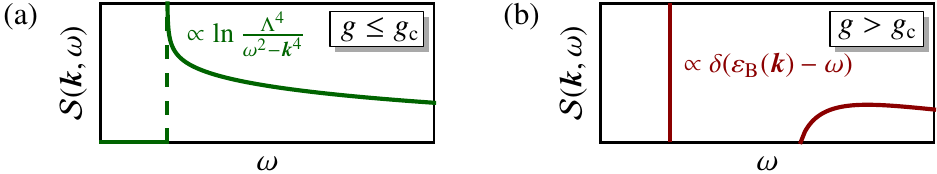}
 \caption{Dynamic structure factor $\mathcal{S}(\vec k, \omega)$ of the order parameter as function of (real) frequency $\omega$ for fixed small wavevector $\vec k$, showing (a)~the critical continuum of excitations for $g \leq g_\mathrm{c}$ and (b)~the quasiparticle peak at the boson excitation energy $\varepsilon_\mathrm{B} (\vec k)$ for~$g > g_\mathrm{c}$.}
\label{fig:spectrum}
\end{figure}

\paragraph{LSM phase.}
%
To reveal the properties of the LSM phase, consider again $\epsilon > 0$ and $g < g_\mathrm{c}$. The low-energy physics is then determined by the LSM fixed point. In the vicinity of this fixed point, the dimensionless mass parameter $r$ (in units of the effective scale $(\Lambda/b)^2$) diverges only slowly in the infrared as $r(b) \propto b^{\epsilon}$. Consequently, the correlation length diverges as $\xi(g) \propto \lim_{b \to \infty} b^{1-\epsilon/2} = \infty$ for all $g$ in the LSM phase.
It is thus a quantum critical phase with emergent scale invariance and power-law correlation functions.
Fig.~\ref{fig:spectrum} shows schematically the dynamic structure factor $\mathcal S(\vec k, \omega)$ of the order-parameter correlations, revealing a continuum of excitations throughout the LSM phase and at the QCP. This is in contrast to the ordered phase, which is characterized by a quasiparticle peak at low energy.

\paragraph{Spin-$1/2$ fermions.}
%
For spin-$1/2$ fermions, the minimal model comprises several four-fermion interactions, leading to a competition between different possible ordered states~\cite{murray2014}. The Gaussian fixed point then realizes a multicritical point. In accordance with the spinless case, correlation functions right at multicriticality display power-law decays, while the order parameters and the correlation length exhibit essential singularities upon approaching the multicritical point~\cite{suppl}.

\paragraph{Discussion.}
%
We have presented an example of an exactly soluble QCP between a stable semimetallic phase with quadratic dispersion and an interaction-induced QAH insulator. On a phenomenological level, this semimetal-insulator transition can be understood as a quantum version of the BKT transition: The semimetallic phase, which corresponds to the low-temperature phase in the BKT transition, is characterized by power-law decays of the correlation functions. In the critical region, the correlation length, the specific heat, and the order parameter exhibit essential singularities. In the insulating phase, the QAH order-parameter field condenses and the correlation length is finite, in analogy to the high-temperature phase of the BKT transition.
For the sake of concreteness, we have chosen for our calculations a particular toy model on the kagome lattice. However, our result of an exactly soluble BKT-like transition applies universally to ordering transitions in clean two-dimensional semimetals with quadratic band touching; remarkably, this even holds independently of the particular symmetry of the ordered phase.
Bernal-stacked bilayer graphene exhibits an ordered ground state below $T_\mathrm{c} \approx 5$\,K~\cite{bao2012}. The fact that the transition temperature $T_\mathrm{c}$ is three orders of magnitude smaller than the characteristic microscopic energy scale~\cite{zhang2008} suggests that remnants of the quantum critical behavior may still be observable in a finite temperature interval above $T_\mathrm{c}$~\cite{ray2018}.
Our predictions for the quantum critical behavior can be directly tested in current numerical simulations of suitable lattice models, including bilayer honeycomb models~\cite{pujari2016}, provided that the interaction strength and the trigonal warping term are simultaneously tuned~\cite{ray2018}.
How to construct a two-dimensional Luttinger semimetal using ultracold atoms has been demonstrated before~\cite{sun2012}, and we anticipate that the QCP proposed here may be experimentally accessible in such a system as well.

\begin{acknowledgments}
We thank I.\ Boettcher, I.\ F.\ Herbut, Z.\ Y.\ Meng, G.\ Plunien, and O.\ Vafek for useful discussions.
This research was supported by the DFG through the Emmy Noether program (JA2306/4-1, project id 411750675), SFB 1143 (project id 247310070), and the W\"urzburg-Dresden Cluster of Excellence \textit{ct.qmat} (EXC 2147, project id 390858490).
\end{acknowledgments}

\end{document}


\title{%
Supplemental Material:\\
Soluble Fermionic Quantum Critical Point in Two Dimensions
}

\author{Shouryya Ray}
\author{Matthias Vojta}
\author{Lukas Janssen}
\affiliation{Institut f\"ur Theoretische Physik, Technische Universit\"at Dresden,
01062 Dresden, Germany}
\affiliation{W\"urzburg-Dresden Cluster of Excellence ct.qmat, Technische Universit\"at Dresden, 01062 Dresden, Germany}


\date{\today}

\maketitle


\section{Low-energy field theory}
This section is devoted to deriving the continuum limit of the Hubbard model on the kagome lattice. The hopping term (tight-binding model), given in Eq.~\eqHtmaintext{} of the main text, is reproduced here for convenience:
\begin{align}
 H_{0} &= -t\sum_{\langle i j \rangle} c_i^\dagger c_j - t' \sum_{\llangle i j \rrangle} c_i^\dagger c_j + \textnormal{H.c.}
\end{align}
As noted in the main text, the low-energy theory is obtained hence by a Schrieffer-Wolff transformation; here, we supply some intermediate steps. In momentum space, the hopping Hamiltonian reads
\begin{align}
 H_{0} &= \int_{\vec{p} \in \text{BZ}}\frac{{d}^2 \vec{p}}{(2\pi)^2}\Psi^\dagger(\vec{p}) \hat{\mathcal{H}}_0(\vec{p}) \Psi(\vec{p}),
\end{align}
where the momentum integration is to be performed over the first Brillouin zone (BZ). The fermion operators on the sublattices A, B, and C are collected into a three-component vector $\Psi^\dagger(\vec{p}) = \left(a^\dagger(\vec{p}),b^\dagger(\vec{p}),c^\dagger(\vec{p})\right)$. The $3\times 3$ matrix $\hat{\mathcal{H}}_0(\vec{p})$ acts in this space and has only off-diagonal entries:
\begin{widetext}
\begin{align}
\hat{\mathcal{H}}_0(\vec{p}) &=
-\left(
\begin{array}{ccc}
 0 & 2t\cos(\vec{\delta}_{\text{AB}}\cdot \vec{p}) + 2t'\cos(\vec{\delta}'_{\text{AB}}\cdot \vec{p}) & 2t\cos(\vec{\delta}_{\text{AC}}\cdot \vec{p}) + 2t'\cos(\vec{\delta}'_{\text{AC}}\cdot \vec{p})\\
 2t\cos(\vec{\delta}_{\text{AB}}\cdot \vec{p}) + 2t'\cos(\vec{\delta}'_{\text{AB}}\cdot \vec{p}) & 0 & 2t\cos(\vec{\delta}_{\text{BC}}\cdot \vec{p}) + 2t'\cos(\vec{\delta}'_{\text{BC}}\cdot \vec{p})\\
 2t\cos(\vec{\delta}_{\text{AC}}\cdot \vec{p}) + 2t'\cos(\vec{\delta}'_{\text{AC}}\cdot \vec{p})& 2t\cos(\vec{\delta}_{\text{BC}}\cdot \vec{p}) + 2t'\cos(\vec{\delta}'_{\text{BC}}\cdot \vec{p}) & 0
\end{array}
\right),
\end{align}
\end{widetext}
where $\vec{\delta}_{\text{AB}}$ ($\vec{\delta}'_{\text{AB}}$) denotes the (next-)nearest neighbor displacement vectors between A and B atoms, and likewise for AC and BC. Explicitly, these may be written as
\begin{align*}
\vec{\delta}_{\text{AB}} &= \tfrac12(1,-\sqrt{3}), \quad \vec{\delta}_{\text{BC}} = \tfrac12(1,\sqrt{3}), \quad \vec{\delta}_{\text{AC}} = (1,0), \\
\vec{\delta}'_{\text{AB}} &= \tfrac12(3,\sqrt{3}), \quad \vec{\delta}'_{\text{BC}} = \tfrac12(3,-\sqrt{3}), \quad \vec{\delta}'_{\text{AC}} = (0,\sqrt{3}),
\end{align*}
with the lattice constant set to unity for notational convenience. Since the quadratic band touching point (QBT) is located at $\vec{p} = 0$, we may expand $\hat{\mathcal{H}}_0(\vec{p})$ for small momenta. To zeroth order, we thus find the eigenvalues
\begin{align}
 {\varepsilon}_{1,2} &= 2t(1 + t'/t) + \mathcal{O}(\vec{p}^2), \\
 {\varepsilon}_{3} &= -4t(1 + t'/t) + \mathcal{O}(\vec{p}^2).
\end{align}
Note that the degenerate pair $ {\varepsilon}_{1,2}$ is merely the manifestation of the QBT at zero momentum. Its eigenspace (henceforth referred to as the low-energy subspace) can be spanned, for instance, by
\begin{align}
  u_1 &= \left(\sqrt{1/6},-\sqrt{2/3},\sqrt{1/6}\right)^\top, \\
  u_2 &= \left(-\sqrt{1/2},0,\sqrt{1/2}\right)^\top.
\end{align}
Given the eigenbasis, we can construct the projector $\mathcal{P}$ onto the low-energy subspace in usual fashion as a sum of dyads, to wit:
\begin{align}
 \mathcal{P} = u_1 u_1^\top + u_2 u_2^\top.
\end{align}
The third eigenvector at $\vec{p} = 0$ is given by
\begin{align}
 u_3 = \left(1/\sqrt{3},1/\sqrt{3},1/\sqrt{3}\right)^\top.
\end{align}
The orthogonal transformation which diagonalizes $\hat{\mathcal{H}}_0(\vec{p}=0)$ can now be constructed from the three eigenvectors as
\begin{align}
 \mathcal{U} = \left(u_1 , u_2 , u_3 \right).
\end{align}
Since the low-energy and the high-energy subspaces are clearly separated, we can finally compute the low-energy sector of $H_0$ at $\mathcal{O}(\vec{p})^2$. It is found \cite{schriefferwolffannphys} by first expanding $\hat{\mathcal{H}}_0(\vec{p})$ to second order in $\vec{p}$, then projecting onto the low-energy subspace using $\mathcal{P}$, and finally block-diagonalizing using $\mathcal{U}$, to wit:
\begin{align}
 H_0 = \int_{\vec{p}} \psi^\dagger(\vec{p}) \mathcal{H}_0(\vec{p}) \psi(\vec{p}) + \cdots\label{eq:HtEFT}
\end{align}
with
\begin{align}
\left(\begin{array}{cc}
\mathcal{H}_0(\vec{p}) & 0 \\
0 & 0
\end{array}\right)
&= \frac12\mathcal{U}^\dagger \mathcal{P} \left[(\partial_\kappa)^2\kern.1em\hat{\mathcal{H}}_0(\kappa\vec{p})\right]_{\kappa \to 0}\mathcal{P} \mathcal{U}.\label{eq:schriefferwolffmaster}
\end{align}
The $\psi(\vec{p})$ are the low-energy 2-spinors, given by
\begin{align}
 \psi_{i} = u_{i}^{\phantom{i}j}\Psi_{j}. \label{eq:lowenergy2spinors}
\end{align}
Note that the terms suppressed in Eq.\ \eqref{eq:HtEFT} are of two kinds: (i) bilinears from the high-energy subspace and (ii) constant energy shifts within the low-energy subspace. Evaluating the above expression [Eq.\ \eqref{eq:schriefferwolffmaster}] for $\mathcal{H}_0(\vec{p})$, we find
\begin{align}
 \mathcal{H}_0(\vec{p}) = \left(
\begin{array}{cc}
 t p_x^2 + 3t' p_y^2 & (t - 3t')p_x p_y\\
 (t - 3t')p_x p_y & 3 t' p_x^2 + t p_y^2
\end{array}
\right).\label{eq:HtEFTeval}
\end{align}
Comparing coefficients, we find that we need $t'/t = -\frac13$ to ensure the particle-hole symmetric QBT Hamiltonian of the low-energy theory considered in the main text [cf.\ Eq.~\eqlowEfourFermi{} therein]. %
To complete the derivation of the low-energy field theory, we need to add the density-density interactions on nearest and next-nearest neighbor bonds [Eq.~\eqHint{} of main text, reproduced here for convenience]:
\begin{align}
 H_{\textnormal{int}} = V_1 \sum_{\langle i j \rangle} c_i^\dagger c_i c_j^\dagger c_j  + V_2 \sum_{\llangle i j \rrangle} c_i^\dagger c_i c_j^\dagger c_j.
\end{align}
To leading order in gradient expansion, these are contact terms, since (second-)nearest neighboring sites belong to different sublattices. We thus find for the low-energy content of $H_{\textnormal{int}}$ in the continuum limit
\begin{align}
 H_{\textnormal{int}} &= 2(V_1 + V_2)\int_{\vec{x}} \left[c_\mathrm{A}^\dagger(\vec{x})\,c_\mathrm{A}(\vec{x})\,c_\mathrm{B}^\dagger(\vec{x})\,c_\mathrm{B}(\vec{x}) \right. \nonumber\\
 &\hphantom{{}={} 2(V_1 + V_2)\int_{\vec{x}} \Big[}\left.{}+ c_\mathrm{A}^\dagger(\vec{x})\,c_\mathrm{A}(\vec{x})\,c_\mathrm{C}^\dagger(\vec{x})\,c_\mathrm{A}(\vec{x}) \right.\nonumber\\
 &\hphantom{{}={} 2(V_1 + V_2)\int_{\vec{x}} \Big[}\left.{}+ c_\mathrm{B}^\dagger(\vec{x})\,c_\mathrm{B}(\vec{x})\,c_\mathrm{C}^\dagger(\vec{x})\,c_\mathrm{C}(\vec{x})\right] + \cdots
\end{align}
where $c_\mathrm{A}^\dagger(\vec x)$ creates a particle in the sublattice A at position $\vec x$, and analogous for the sublattices B and C.
Writing out in terms of the low-energy 2-spinor $\psi = (\psi_1,\psi_2)$ using Eq.\ \eqref{eq:lowenergy2spinors}, we hence find
\begin{align}
 H_{\text{int}} %
 &= 2(V_1 + V_2) \int_{\vec{x}} \psi_1^\dagger(\vec{x}) \psi_1(\vec{x}) \psi_2^\dagger(\vec{x}) \psi_2(\vec{x}) \\
 &= -2(V_1 + V_2)\int_{\vec{x}} \tfrac12\left(\psi^\dagger(\vec{x})\sigma_2\psi(\vec{x})\right)^{\kern-.1em 2}.
\end{align}
For the low-energy field theory, we require that the coefficient of the kinetic term be unity. The most efficient way to achieve this is at the level of the action, by noting that the relevant contributions have the form $S \supset \int{d}\tau \left(H_0 + H_{\textnormal{int}}\right)$; rescaling $t\tau \to \tau$ thus absorbs all coefficients of $t$ appearing in $\mathcal{H}_0(\vec{p})$ (we restrict ourselves to $t > 0$). %
Concomitantly, this leads to $H_{\text{int}} \to H_{\text{int}}/t$, whence comparison of coefficients for the four-fermion term yields the identification $g = 2(V_1 + V_2)/t$, just as we asserted in the main text.

\section{Particle-hole asymmetry}
\begin{figure}
 \includegraphics[scale=1]{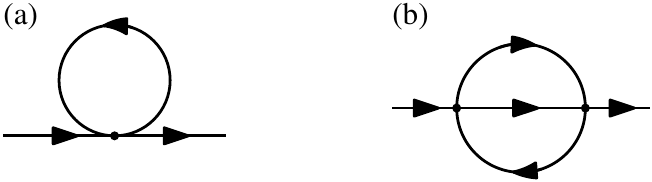}
 \caption{Self-energy corrections at one-loop (a) and two-loop (b) order. In the fermionic model, the Hartree diagram (a) vanishes, while the sunset diagram (b) leads to the contribution displayed in Eq.~\eqref{eq:sunset}.}
 \label{fig:ph-asymmetry}
\end{figure}
In the main text, we have assumed a model for a particle-hole-symmetric point of QBT. Here, we comment on the effect of small symmetry-breaking perturbations. On the level of the continuum field theory, the only term compatible with the remaining symmetries---rotational invariance and time-reversal---that is not power-counting irrelevant corresponds to a difference in the curvatures of the two bands that touch at the QBT~\cite{sun2009},
\begin{align}
	\mathcal L_{\alpha} = - \alpha \psi^\dagger \left(\partial_x^2 + \partial_y^2 \right) \psi,
\end{align}
with dimensionless parameter $\alpha$. The flow of $\alpha$ is given by the selfenergy; however, in the fermionic model, the leading-order (Hartree) diagram [Fig.~\ref{fig:ph-asymmetry}(a)] vanishes and $\alpha$ hence does not flow at the one-loop order. To decide the fate of small particle-hole asymmetric perturbations, we thus have to evaluate the two-loop diagram, Fig.~\ref{fig:ph-asymmetry}(b). This diagram has a sunset topology, which allows the loop integration to be performed efficiently by going to \emph{position space} and back \cite{groote99,ray2018}, during the course of which the two coupled integrals over 3-momenta essentially decompose into two independent Fourier transforms.
The contribution of the sunset diagram is given by
\begin{align}
\text{Fig.~\ref{fig:ph-asymmetry}(b)} &= g^2\int d\tau\,d^2\vec{x}\,\rme^{-\rmi (\omega \tau + \vec{p}\cdot \vec{x})} \sigma_2\kern.1em\tilde{G}_0(\tau,\vec{x})\kern.1em\sigma_2 \times {} \nonumber\\
&\phantom{{}\sim{}}{}\times (\operatorname{tr} - 1)\left[\tilde{G}_0(-\tau,-\vec{x})\kern.1em\sigma_2\kern.1em\tilde{G}_0(\tau,\vec{x})\kern.1em\sigma_2\right],
\label{eq:sunset}
\end{align}
where
\begin{align}
\tilde{G}_0(\tau,\vec{x}) = \int \frac{d\omega\, d^2\vec{p}}{(2\pi)^3}\,\frac{\rme^{\rmi (\omega \tau + \vec{p}\cdot \vec{x})}}{\rmi \omega + d_a(\vec{p})\kern.1em\sigma_a + \alpha \vec{p}^2}
\end{align}
is the bare fermion propagator in position space, with $\sigma_a = (\sigma_3, \sigma_1)$ and $d_a(\vec{p}) = \bigl(p_x^2 - p_y^2,2p_x p_y\bigr)$. For small particle-hole asymmetry $\alpha$, all expressions in general, and the propagator in particular, may be expanded in powers of $\alpha$:
\begin{align}
 G_0(\omega,\vec{p}) &= \frac{-\rmi\omega + d_a(\vec{p})\kern.1em\sigma_a}{\omega^2 + \vec{p}^4} \nonumber\\
 &\phantom{{}={}} {} + \alpha\,\frac{\vec{p}^2\left(\omega^2 - \vec{p}^4\right) + 2\rmi \omega \vec{p}^2 d_a(\vec{p})\kern.1em\sigma_a}{\left(\omega^2 + \vec{p}^4\right)^2} + \mathcal{O}(\alpha^2).
\end{align}
The integral over $\omega$ is elementary. For the spatial part of the Fourier transform, introduce cylindrical coordinates $\vec{p} = p(\cos\varphi,\sin\varphi)$, $\vec{x} = \varrho(\cos\vartheta,\sin\vartheta)$. Then, $\vec{p}\cdot \vec{x} = p\varrho \cos(\varphi - \vartheta)$, and subsequently
\begin{align}
 \rme^{\rmi \vec{p}\cdot \vec{x}} &= J_0(p\varrho) + 2\sum_{m=0}^{\infty}\rmi^m J_m(p\varrho)\left[\cos(m\varphi)\cos(m\vartheta) \right. \nonumber\\
 &\hphantom{{}={}J_0(p\varrho) + 2\sum_{m=0}} {} + \left.\sin(m\varphi)\sin(m\vartheta)\right]
\end{align}
by the Jacobi-Anger identity \cite{abrasteg}, where $J_m(\cdot)$ is the Bessel function of the first kind and order $m$. The integral over $\varphi$ can now be performed by exploiting the orthogonality of sines and cosines in $L^2([0,2\pi])$, and the series conveniently terminates at $m = 2$ due to rotational invariance. The final integral over $\varrho$ turns out to be expressible in terms of elementary functions as well, yielding explicit expressions for the tree-level position-space propagator to the desired order in $\alpha$:
\begin{align}
 \tilde{G}_0(\tau,\vec{x}) &= \frac{\rme^{-\vec{x}^2/(4|\tau|)}}{8\pi\tau} + \frac{\rme^{-\vec{x}^2/(4|\tau|)}\left(\vec{x}^2 + 4|\tau|\right) - 4|\tau|}{8\pi \vec{x}^2 |\tau|}\frac{d_a(\vec{x})}{\vec{x}^2}\sigma_a \nonumber\\
 &\phantom{{}={}} {}+ \frac{\rme^{-\vec{x}^2/(4|\tau|)}}{32\pi \tau^2} \left[\vec{x}^2 - 4|\tau| + \operatorname{sgn}(\tau)\, d_a(\vec{x})\kern.1em\sigma_a\right]\alpha \nonumber\\
 &\phantom{{}={}} {}+ \mathcal{O}(\alpha^2).
\end{align}
To obtain the selfenergy in momentum space, we have to perform the inverse Fourier transform, followed by an expansion in powers of external momentum $\vec{p}$ to extract renormalization constants. The integral is both infrared and ultraviolet divergent, so we perform a regularization using sharp cutoffs $1/\Lambda < |\vec{x}| < b/\Lambda$. (The integration over $\omega$ requires no further regularization.) Conceptually, this is equivalent to Wilsonian renormalization, except that the degrees of freedom being integrated out are supported on a shell in \textit{position} (rather than momentum) space; the notation for the cutoffs is chosen to make this analogy (cf.\ Sec.~\ref{sec:RG}) transparent. We thus find
\begin{align}
\text{Fig.~\ref{fig:ph-asymmetry}(b)} &= \frac{g^2}{4\pi^2}\left[\frac{1}{24}d_a(\vec{p})\kern.1em\sigma_a + \alpha\left(\frac19 - \frac14 \ln \frac43\right)\vec{p}^2\right]\ln b \nonumber \\ &= \eta_\psi d_a(\vec{p})\kern.1em\sigma_a\ln b + \delta \alpha \vec{p}^2,
\end{align}
whence
\begin{align}
\frac{d \alpha}{d\ln b} = \frac{\partial\kern.1em\delta \alpha}{\partial \ln b} - \eta_\psi \alpha = -\frac{g^2}{4\pi^2}\left(\frac14\ln\frac43 - \frac{5}{72}\right)\alpha.
\end{align}
Importantly, the derivative of the right-hand side with respect to $\alpha$ is negative, and shows that small particle-hole asymmetry is an irrelevant perturbation.

\section{Renormalization group equations for the Luttinger-Yukawa theory}
\label{sec:RG}
In this section, we provide some details on the partially bosonized formulation of the four-fermion theory. We restrict ourselves here to the technical computations needed to derive the anomalous dimensions and $\beta$-functions and the fixed points of the RG flow; their physical content is discussed in the main text. For ease of reference, we recall here the Lagrange density, which is given by
\begin{align}
\mathcal{L}' & =
  \psi^\dagger\left[\partial_\tau + d_a(-\rmi \nabla)\kern.1em\sigma_a \right]\psi
  + \frac12 \phi \left(r - c \partial_\tau^2 - \partial_x^2 - \partial_y^2\right) \phi
  \nonumber \\ & \quad
   - h \phi \psi^\dagger \sigma_2 \psi,
   \label{eq:Lprime}
\end{align}
where $\sigma_a = (\sigma_3, \sigma_1)$ and $d_a(\vec{p}) = (p_x^2 - p_y^2,2p_x p_y)$. The equation of motion for $\phi$ yields $\phi = (h/r)\psi^\dagger \sigma_2 \psi$, and we need to identify $g = h^2/r$ to connect $\mathcal{L}'$ to the four-fermion Lagrangian. The Hubbard-Stratonovich transformation by itself yields no dynamical terms for $\phi$, but they would be generated under RG flow anyway, and might as well be included from the outset. The canonical dimensions can now be read off in usual manner; they are given by $[\psi] = [\phi] = (d + z - 2)/2$, $[r] = 2$, $[c]= 2 - 2z$, and $[h^2] = 6 - d - z$, with $z = 2$ at tree level.

To find the RG flow of these quantities, we perform a loop expansion. Since we are only interested in one-loop results, the Wilsonian scheme is particularly well-suited, as it efficiently regulates IR divergences which would otherwise appear due to the presence of massless internal fermion lines. To make the exposition self-contained (but also to fix some notation), we recapitulate the operational details of Wilsonian RG in the present context. We assume an action $S_\Lambda[\Phi_\Lambda,X_\Lambda]$ with sharp UV cutoff $\Lambda$, where we have collected all fields and couplings in the theory into $\Phi = (\psi,\psi^\dagger,\phi)$ and $X = (c,r,h)$ (when no explicit scale is mentioned, a relation is meant to be valid at all scales). We next integrate out all momenta in the shell $|\vec{p}| \in [{\Lambda/b},\Lambda]$, $b > 1$ (note that the integration over imaginary loop frequencies is unrestricted); the corresponding one-particle irreducible (1PI) effective action is denoted $\Gamma_{\Lambda,{\Lambda/b}}[\Phi_\Lambda,X_\Lambda]$. To impose RG invariance is to demand that this change in the action be equivalent to a suitable redefinition of fields and couplings, to wit: %
\begin{align}
  \Gamma_{\Lambda,{\Lambda/b}}[\Phi_\Lambda,X_\Lambda] = S_\Lambda[\Phi_{{\Lambda/b}},X_{{\Lambda/b}}].
\end{align}
\begin{widetext}
We shall perform an infinitesimal RG step, i.e., $b = 1 + \ln b+ \mathcal{O}\!\left((\ln b)^2\right)$; momentum integrals then simplify to
\begin{align}
 \int_{|\vec{p}| \in [{\Lambda/b},\Lambda]} d^d \vec{p}\,f(\vec{p}) = \Lambda^d \ln b \int_0^{2\pi}{d} \varphi_p\,f(\Lambda \cos\varphi_p,\Lambda \sin\varphi_p) + o\!\left(\ln b\right).
\end{align}
The above equation assumes a minimal prescription for analytic continuation to arbitrary spatial dimension $d$: only the radial integration is modified, while angular integration and spinor algebra is left at $d = 2$. Since we do not intend to eventually make predictions for nonzero $(d - 2)$, this is guaranteed to be equivalent to more sophisticated schemes for our purposes.

At tree level, the UV modes do not contribute at all, and $\Gamma_{\Lambda,{\Lambda/b}}[\Phi_\Lambda,X_\Lambda] = S_{{\Lambda/b}}[\Phi_\Lambda,X_\Lambda]$. The one-loop diagrams are shown in Fig.\ \ref{fig:1Loop}, corrections arising from which have the schematic form
\begin{align}
 &\Gamma_{\Lambda,{\Lambda/b}}[\Phi_\Lambda,X_\Lambda] - S_{{\Lambda/b}}[\Phi_\Lambda,X_\Lambda] \nonumber\\
 &\quad {}= \int \frac{{d}\omega\,{d}^d \vec{p}}{(2\pi)^{d+1}}\Theta({\Lambda/b} - |\vec{p}|)\left[\hat{\psi}^\dagger_0(\omega,\vec{p})\left(\delta Z_\psi d_a(\vec{p}) \sigma_a + \delta Z_\omega\rmi \omega\right)\hat{\psi}_0(\omega,\vec{p})
  + \tfrac12 \hat{\phi}_0\vphantom{{}^\dagger}(\omega,\vec{p})\left(\delta Z_\phi \vec{p}^2 + \delta c\,\omega^2 + \delta r\right)\hat{\phi}_0(-\omega,-\vec{p})\right] \nonumber\\
 &\quad \hphantom{{}={}}{} - \int\left({\textstyle\prod_{i}} \frac{{d}\omega_i{d}^d\vec{p}_i}{(2\pi)^{d+1}}\Theta({\Lambda/b} - |\vec{p}_i|)\right)\delta\!\left({\textstyle\sum_{i}}\omega_i\right)\delta^{(d)}\!\left({\textstyle\sum_i} \vec{p}_i\right)\delta h\,\hat{\phi}(\omega_1,\vec{p}_1)\,\hat{\psi}^\dagger(-\omega_2,-\vec{p}_2)\kern.1em\sigma_2\,\hat{\psi}(\omega_3,\vec{p}_3) \nonumber\\
 & \quad \hphantom{{}={}}{} - \int\left({\textstyle\prod_{i}} \frac{{d}\omega_i{d}^d\vec{p}_i}{(2\pi)^{d+1}}\Theta({\Lambda/b} - |\vec{p}_i|)\right)\delta\!\left({\textstyle\sum_{i}}\omega_i\right)\delta^{(d)}\!\left({\textstyle\sum_i} \vec{p}_i\right)\tfrac12 \delta g\,\hat{\psi}^\dagger(-\omega_1,-\vec{p}_1)\kern.1em\sigma_2\,\hat{\psi}(\omega_2,\vec{p}_2)\,\hat{\psi}^\dagger(-\omega_3,-\vec{p}_3)\kern.1em\sigma_2\,\hat{\psi}(\omega_4,\vec{p}_4).
\end{align}
Here, the ``hat'' on a field is used to denote its Fourier transform: $\hat{\Phi}(\omega,\vec{p}) = \int {d}\tau{d}^d\vec{x}\,\rme^{-\rmi (\omega \tau + \vec{p}\cdot \vec{x})}\Phi(\tau,\vec{x})$. %
Note that $[\hat{\Phi}] = [{d}\tau{d}^d\vec{x}\,\Phi] = [\Phi] - d - z$. The necessary coupling and scale redefinitions can now be read off. In the same schematic manner, they are given by
\begin{align}
  z &= 2 - \eta_\psi + \eta_\omega, \label{eq:zdef}\\
  \psi_{{\Lambda/b}} &= \psi_\Lambda\left[1 + \tfrac12 (d + z - 2 + \eta_\psi)\ln b\right], \\
  \phi_{{\Lambda/b}} &= \phi_\Lambda\left[1 + \tfrac12 (d + z - 2 + \eta_\phi)\ln b\right] - (2h_\Lambda)^{-1}\ln b\left(\partial\delta g/\partial \ln b\right)_{b \to 1}\psi_\Lambda^\dagger \sigma_2 \psi_\Lambda,\\
  c_{{\Lambda/b}} &= c_\Lambda\left[1 + (2 - 2z - \eta_\phi)\ln b\right] + \ln b\left(\partial\delta c/\partial \ln b\right)_{b \to 1} \equiv c_\Lambda + \beta_c \ln b,\\
  r_{{\Lambda/b}} &= r_\Lambda\left[1 + (2 - \eta_\phi)\ln b\right] + \ln b\left[\partial\delta r/\partial \ln b\right]_{b \to 1} \equiv r_{{\Lambda/b}} + \beta_r \ln b, \\
  h_{{\Lambda/b}} &= h_\Lambda\left[1 + \tfrac12 (6 - d - z - \eta_\phi - 2\eta_\psi)\ln b\right] + \left\{\vphantom{(2h_\Lambda)^{-1}}\left(\partial\delta h/\partial \ln b\right)_{b \to 1} \right.\nonumber \\
  &\qquad\left. {} + r_\Lambda(2h_\Lambda)^{-1}\left(\partial\delta g/\partial \ln b\right)_{b \to 1}\right\}\ln b \equiv h_\Lambda + (2h_\Lambda)^{-1}\beta_{h^2}\ln b,
  \label{eq:betah2def}
\end{align}
where $\eta_{\psi,\phi,\omega} = (\partial \delta Z_{\psi,\phi,\omega} / \partial \ln b)_{b \to 1}$, and we have introduced the $\beta$-functions $\beta_X = \partial X_{\Lambda/b}/\partial \ln b$. Note that the field renormalization of $\phi$ has a nonmultiplicative component. It is needed to rewrite the four-fermion vertex (which is marginal at $d = 2$ and $z = 2$) generated by the so-called ``box diagrams'' as a renormalization of the Yukawa vertex (see Refs.~\cite{wetzel,luperini} for the analogous demonstration in the relativistic $(1+1)$D Gross-Neveu theory). One may think of it as performing a Hubbard-Stratonovich transformation \emph{after} each RG step, which is why this procedure is also referred to as ``dynamical bosonization'' \cite{dynamrebos} (particularly in the context of the functional renormalization group).
\begin{figure*}
 \includegraphics[width=\textwidth]{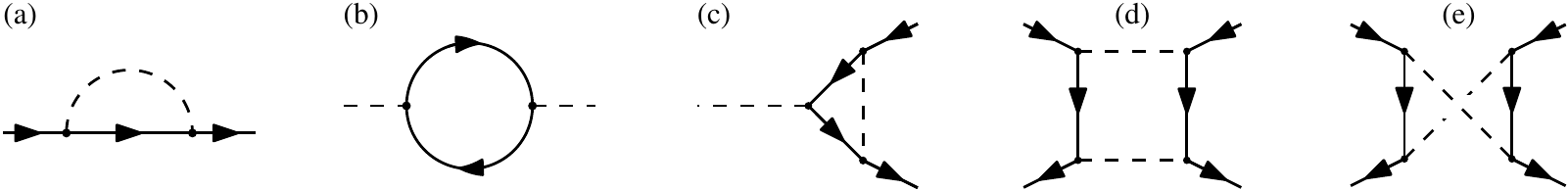}
 \caption{Relevant [(a)--(c)] and marginal [(d), (e)] contributions to the effective action at 1-loop order for the theory Luttinger-Yukawa theory. The vertex is $h \sigma_2$; the solid and dashed lines represent fermions and $\phi$-bosons respectively.}
 \label{fig:1Loop}
\end{figure*}

We now evaluate the diagrams explicitly (external 3-momentum is always $(\omega,\vec{p})$ unless zero; we drop the RG scale index for brevity; $G_0$ is the fermion propagator and $D_0$ is the boson propagator):

\paragraph*{Fermion selfenergy.}
\begin{align}
 \textnormal{Fig.\ \ref{fig:1Loop}(a)} &= -h^2\int_{|\vec{p}^\prime|\in [{\Lambda/b},\Lambda]}\frac{{d} \omega^\prime\,{d}^d\vec{p}^\prime}{(2\pi)^{d+1}}\,\sigma_2 G_0(\omega^\prime,\vec{p}^\prime)\sigma_2 D_0(\omega-\omega^\prime,\vec{p}-\vec{p}^\prime) \nonumber\\
 &= \frac{h^2}{\Lambda^{4-d}(2\pi)^{d}}\left[\frac{2\pi c\Lambda^2}{\left(1 + r/\Lambda^2 - c\Lambda^2\right)^2}\rmi\omega + \frac{\pi}{\left(1 + r/\Lambda^2 - c\Lambda^2\right)^3}d_a(\vec{p})\kern.1em\sigma_a\right]\ln b \nonumber \\
 &\equiv \delta Z_\omega\,\rmi\omega + \delta Z_\psi\,d_a(\vec{p})\kern.1em\sigma_a
\label{eq:fermionselfenergy}
\end{align}

\paragraph*{Boson vacuum polarization.}
\begin{align}
 \textnormal{Fig.\ \ref{fig:1Loop}(b)} &= h^2\int_{|\vec{p}^\prime|\in [{\Lambda/b},\Lambda]}\frac{{d} \omega^\prime\,{d}^d\vec{p}^\prime}{(2\pi)^{d+1}}\,\operatorname{tr}\!\left[G_0(\omega^\prime,\vec{p}^\prime)\sigma_2 G_0(\omega^\prime-\omega,\vec{p}^\prime - \vec{p})\sigma_2\right] \nonumber\\
 &= \frac{h^2}{\Lambda^{4-d}(2\pi)^d}\left(\frac{\pi}{2\Lambda^2}\omega^2 + \pi\vec{p}^2 - 2\pi\Lambda^2\right)\ln b \nonumber\\
 &\equiv \delta c\,\omega^2 + \delta Z_\phi\,\vec{p}^2 + \delta r
\end{align}

\paragraph*{Yukawa vertex correction.}
\begin{align}
 \textnormal{Fig.\ \ref{fig:1Loop}(c)} &= -h^3 \int_{|\vec{p}^\prime|\in [{\Lambda/b},\Lambda]}\frac{{d} \omega^\prime\,{d}^d\vec{p}^\prime}{(2\pi)^{d+1}}\,\sigma_2 G_0(\omega^\prime,\vec{p}^\prime) \sigma_2 G_0(\omega^\prime,\vec{p}^\prime) \sigma_2 D_0(\omega^\prime,\vec{p}^\prime) \nonumber\\
 &= \frac{h^3}{\Lambda^{4-d}(2\pi)^d}\sigma_2\frac{\pi}{1 + r/\Lambda^2 + \sqrt{c\Lambda^2(1+r/\Lambda^2)}}\ln b \nonumber\\
 &\equiv -\delta h\,\sigma_2
\end{align}

\paragraph*{Four-fermion vertex (nonmultiplicative renormalization of $\phi$).}
\begin{align}
 \textnormal{Fig.\ \ref{fig:1Loop}(d, e)} &= - h^4 \int_{|\vec{p}^\prime|\in [{\Lambda/b},\Lambda]}\frac{{d} \omega^\prime\,{d}^d\vec{p}^\prime}{(2\pi)^{d+1}}\,\sigma_2 G_0(\omega^\prime,\vec{p}^\prime) \sigma_2 D_0(\omega^\prime,\vec{p}^\prime) \otimes \sigma_2 \left[G_0(\omega^\prime,\vec{p}^\prime) + G_0(-\omega^\prime,-\vec{p}^\prime)\right] \sigma_2 D_0(\omega^\prime,\vec{p}^\prime) \nonumber \\
 &= - \frac{h^4\pi}{\Lambda^{6-d}(2\pi)^d}\frac{\left(1+r/\Lambda^2\right)^{5/2} - 5c\Lambda^2\left(1+r/\Lambda^2\right)^{3/2} + 5c^{3/2}\Lambda^3\left(1+r/\Lambda^2\right) - c^{5/2}\Lambda^5}{\left(1+r/\Lambda^2\right)^{3/2}\left(1 + r/\Lambda^2 - c\Lambda^2\right)^3}\sigma_2 \otimes \sigma_2\,\ln b\nonumber \\
 &\equiv - \delta g\,\sigma_2 \otimes \sigma_2.
 \label{eq:4Fermivertex}
\end{align}
Note that we have made use of the Fierz identity $(\psi^\dagger \sigma_a \psi)^2 = 2(\psi^\dagger \sigma_2 \psi)^2$ in the second equality to convert $\sigma_a \otimes \sigma_a$ to $\sigma_2 \otimes \sigma_2$.
\end{widetext}

Eqs.~\eqref{eq:fermionselfenergy}--\eqref{eq:4Fermivertex} fix the renormalization constants, and inserting these expressions into Eqs.~\eqref{eq:zdef}--\eqref{eq:betah2def} yield the RG flow equations. To simplify the notation, we rescale %
$r/\Lambda^2 \mapsto r$, %
$c \Lambda^2 \mapsto c$, %
and %
$(2\pi)^{-\epsilon}\Lambda^{-(2+\epsilon)} h^2 \mapsto h^2$, where $\epsilon = d - 2$. The $\beta$-functions then read as:
\begin{align}
  \beta_r & = \bigl(2 - \eta_\phi\bigr) r - \frac{h^2}{2\pi}, \\
  \beta_c & = \bigl(- 2 - \eta_\phi\bigr) c + \frac{h^2}{8\pi}, \\
  \beta_{h^2} &= \bigl(2 - \epsilon - \eta_\phi - 2\eta_\psi\bigr)h^2 - \frac{h^4}{4\pi}\left[\frac{2}{1+r+ c^{1/2}(1+r)^{1/2}} \right. \nonumber\\
  &\phantom{{}={}}\left.{}- \frac{(1+r)^{5/2} - 5c(1+r)^{3/2} + 5c^{3/2}(1+r) - c^{5/2}}{(1+r)^{3/2}(1+r-c)^3}\right] \nonumber\\
  &= \bigl(2 - \epsilon - \eta_\phi\bigr)h^2 - \frac{h^4}{4\pi r} + \mathcal{O}(1/r^2),
\end{align}
with the anomalous dimensions $\eta_{\phi,\psi}$ and dynamical critical exponent $z$ given by
\begin{align}
  z &= 2 - \frac{h^2}{4\pi}\left[\frac{1}{(1+r-c)^3}-\frac{2c}{(1+r-c)^2}\right] \nonumber\\
  &= 2 + \mathcal{O}(1/r^2), \\
  \eta_\psi &= \frac{h^2}{4\pi}\frac{1}{(1+r-c)^3} = \mathcal{O}(1/r^3), \\
  \eta_\phi &= \frac{h^2}{4\pi}.
  \label{eq:etaphiraw}
\end{align}
Note that we have expanded all expressions in powers of $1/r$, in anticipation of the fact that all pertinent fixed points will be characterized by a large fixed-point value of $r$, at least of order $1/\epsilon$ (such that $h^2/r = g \ll 1$). As a cross-check, it is useful to compute entirely within this theory the $\beta$-function of the effective four-fermion coupling $g = h^2/r$. We find%
\begin{align}
 \beta_{h^2/r} = \frac{\beta_{h^2}}{r} - \frac{h^2 \beta_r}{r^2} = -\epsilon \frac{h^2}{r} + \frac{1}{4\pi}\left(\frac{h^2}{r}\right)^2,
\end{align}
which agrees precisely with the result obtained in the four-fermion theory.

For $\epsilon > 0$, the RG flow admits two nontrivial fixed points, located (to first order in $\epsilon$) at:
\begin{align}
 \text{LSM} &\colon %
 \left(r^\star_\text{LSM}, c^\star_\text{LSM}, h^{\star2}_\text{LSM}\right) = \left(\infty, \frac14 - \frac{1}{16}\epsilon, 4\pi(2 - \epsilon)\right),
 \label{eq:LSMFP}
 \\
 \text{QCP} &\colon %
 \left(r^\star_\text{QCP}, c^\star_\text{QCP}, h^{\star2}_\text{QCP}\right) = \left( \frac{2}{\epsilon}, \frac14-\frac18\epsilon, 8\pi(1-\epsilon)\right).
 \label{eq:QCPFP}
\end{align}
The second fixed point has precisely one relevant direction, and hence governs the semimetal-insulator transition; this justifies the label QCP. Inserting $h^{\star2}_\text{QCP}$ into Eq.~\eqref{eq:etaphiraw} yields the (fixed-point value of) the order-parameter anomalous dimension at the QCP, $\eta_{\phi}|_\text{QCP} = 2 - 2\epsilon$. To obtain the correlation-length exponent, one needs to study the eigenvalues of the stability matrix
\[
 \left.\left(\partial_{X}\beta_{X'}\right)\right|_{X,X'=\left(r^\star_\text{QCP}, c^\star_\text{QCP}, h^{\star2}_\text{QCP}\right)}
\]
at the QCP. There are three distinct eigenvalues, the only positive one among which reads as $\lambda_+ = \epsilon + \mathcal{O}(\epsilon^2)$. Hence we find $\nu = 1/\lambda_+ = 1/\epsilon + \mathcal{O}(\epsilon^0)$, which agrees with---but is completely independent of---the calculation in the four-fermion formulation, thus furnishing another nontrivial consistency check. The remaining critical exponents can be obtained from $\eta_{\phi}|_\text{QCP}$ and $\nu$ using hyperscaling.

The other nontrivial fixed point, Eq.~\eqref{eq:LSMFP}, is the only infrared stable (i.e., stability matrix has no positive eigenvalues) fixed point in this theory. It is hence the only candidate for the Luttinger-Yukawa incarnation of the Gaussian fixed point of the four-fermion theory, and describes the stable Luttinger semimetal phase in the bosonized formulation.

\section{Effective potential}
Here we derive the effective potential quoted in the main text [cf.\ Eq.~\eqeffpotmaintext{} therein]. To this end, we start from the Yukawa theory with Lagrangian $\mathcal{L}'$. At tree level, the effective potential is the classical one, $V_{\text{eff},0}(\phi) = r\phi^2/2$. The leading quantum correction $V_{\text{eff},1}(\phi)$ comes from integrating out the fermions in the presence of a constant $\phi$. The result is given by the usual ``trace-log'' formula. Technically, it is simpler to differentiate with respect to $\phi$ first, yielding
\begin{align}
V_{\text{eff},1}'(\phi) = \int_{|\vec{p}| \leq \Lambda} \frac{d\omega\,d^2\vec{p}}{(2\pi)^3}\operatorname{tr}\frac{h\sigma_2}{\rmi \omega + d_a(\vec{p})\kern.1em\sigma_a - h\phi\sigma_2}.
\end{align}
A UV divergence occurs when integrating over spatial 2-momenta, which we have regularized with a sharp cutoff $\Lambda$. The integral over Euclidean frequencies is finite (at least in the sense of a Cauchy principal value), and needs no further regularization. This is precisely (the UV part of) the regularization scheme we used when computing the RG flow in Sec.~\ref{sec:RG}; the regularization of the IR divergence is automatically implemented by the finite background field $\phi$. The loop integral is now straightforward, and the result (after integrating with respect to $\phi$) reads
\begin{align}
V_{\text{eff},1}(\phi) = \frac{h^2\phi^2}{16\pi}\left(\ln\frac{h^2\phi^2}{4\Lambda^4} - 1\right),
\label{eq:Veff1L}
\end{align}
where we have neglected all terms which vanish for $\Lambda \to \infty$. In the limit of a large number of fermion flavors $N$, the only surviving loop contribution is $NV_{\text{eff},1}(\phi)$. It is also mean-field in the sense that fluctuations of the boson field $\phi$ do not enter at any stage (since the fermions were integrated out for constant field configurations of $\phi$). The present case, $N = 1$, is in some sense the opposite limit, and fluctuations of $\phi$ are particularly important. To take them into account, we employ the Callan-Symanzik equation \cite{QFTbook}, which expresses the RG invariance of the effective potential and reads
\begin{align}
\left(\Lambda\frac{\partial}{\partial \Lambda} - \gamma_{h^2}h^2\frac{\partial}{\partial h^2} - \gamma_{r}r\frac{\partial}{\partial r} - \frac12 \eta_\phi \phi\frac{\partial}{\partial \phi}\right)V_{\text{eff}}(\phi) = 0.
\label{eq:callansymanzik}
\end{align}
Here, the quantity $\gamma_X = \beta_X/X - [X]$ describes the ``anomalous'' scaling of the coupling $X$, while $\eta_\phi$ is the anomalous dimension of the field $\phi$ as before. In the limit of small four-fermion coupling $h^2/r \ll 1$, we have
\begin{align}
\eta_\phi &= \frac{h^2}{4\pi \Lambda^2} \equiv \eta_\phi^{\text{MF}}, \\
\gamma_{r} &= -\frac{h^2}{4\pi \Lambda^2} - \frac{h^2}{2\pi r} \equiv \gamma_{r}^{\text{MF}}, \\
\gamma_{h^2} &= -\frac{h^2}{4\pi \Lambda^2} - \frac{h^2}{4\pi r} \equiv \gamma_{h^2}^{\text{MF}} - \frac{h^2}{4\pi r}.
\label{eq:gamma1L}
\end{align}
For future reference, we have split the contributions further into two parts: those coming from (i) diagrams without internal $\phi$-lines (hence mean-field, ``MF''), i.e., the vacuum polarization diagram Fig.~\ref{fig:1Loop}(a), and (ii) diagrams with virtual $\phi$-bosons, which would not survive in the mean-field limit. The latter concerns the last term in $\gamma_{h^2}$, and arises due to the triangle diagram [Fig.~\ref{fig:1Loop}(c)] and the bosonization of the box diagrams [Figs.~\ref{fig:1Loop}(d) and (e)].

Let us now recall the structure of the full effective potential to all loop orders, which is given by
\begin{align}
V_{\text{eff}}(\phi) = \frac{h^2\phi^2}{2h^2/r}\sum_{n=0}^{\infty}\sum_{m \leq n} C_{n,m} \left(\frac{1}{16\pi}\frac{h^2}{r}\right)^n\left(\ln\frac{h^2\phi^2}{4\Lambda^4}\right)^m,\label{eq:Veffallloops}
\end{align}
as a consequence of Collins' theorem. Formally, the contributions at fixed $n$ arise from $n$-loop vacuum diagrams. However, even to leading order in $h^2/r$, observables such as the vacuum expectation value $\langle \phi \rangle$ are sensitive to so-called ``leading logarithms'' (terms with $m = n$ in the series above); these contributions (including multiloop ones, i.e., $n \geq 2$) are fixed \emph{entirely} by the one-loop RG functions (we refer again to textbooks such as \cite{QFTbook}). We can hence compute the coefficients $C_{n,n}$ for all $n$ by inserting the ansatz, Eq.~\eqref{eq:Veffallloops}, into the Callan-Symanzik equation [Eq.~\eqref{eq:callansymanzik}], with the 1-loop RG functions given by Eq.~\eqref{eq:gamma1L}.

At one loop ($n = 1$), we obtain $C_{1,1} = 2$; as a byproduct, we also find $C_{1,0} = -2$. This is consistent with the explicit calculation, Eq.~\eqref{eq:Veff1L}, and expresses the fact that the one-loop effective potential is completely determined by the one-loop RG functions. Note that the same result would have been found if we had used the mean-field expressions for the RG functions, which is reassuring. For $n = 2$, we subsequently find $C_{2,2} = C_{1,1}/2 = 1$, which goes beyond the mean-field level. Remarkably, all further leading logarithms are found to vanish: $C_{n,n} = 0$ for all $n \geq 3$. Thus, the effective potential given by
\begin{align}
  V_{\text{eff}}(\phi)
  = \frac{h^2\phi^2}{2h^2/r}\left[1 + \frac{h^2/r}{8\pi}\left(\ln\frac{h^2\phi^2}{4\Lambda^4}-1\right)
+ \left(\frac{h^2/r}{16 \pi}\ln\frac{h^2\phi^2}{4\Lambda^4}\right)^2\right]
\end{align}
includes leading logarithms to all loops and subleading logarithms to one loop; it is the result quoted in Eq.~\eqeffpotmaintext{} of the main text. We note in passing, that the ``RG-improved'' classical potential reproduces the vanishing of leading logarithms at three-loop and higher order. The quantity we need to compute for this purpose is
\begin{align}
V_{\text{eff}}^{\text{LL}}(\phi) = \frac{h^2(t)\phi^2(t)}{2 (h^2/r)(t)},
\end{align}
where $t = \ln(\Lambda_0/\Lambda)$ refers to the RG time (we have reinterpreted $\Lambda$ as the running scale and the UV cutoff is now denoted $\Lambda_0$). After performing the necessary integrations, one has
\begin{align}
V_{\text{eff}}^{\text{LL}}(\phi) \propto \left[(h^2/r)\cdot t - 4\pi\right]^2,
\end{align}
where couplings without RG time argument refer to the respective initial values at the UV scale (we have suppressed $t$-independent factors). For the resummation of leading logarithms to take effect, we have to choose the running scale $\Lambda$ so that the logarithms in Eq.~\eqref{eq:Veffallloops} vanish, to wit: $\Lambda^4 = h^2\phi^2/4$, whence $t = -\frac14\ln\left(h^2\phi^2/4\Lambda_0^4\right)$. Importantly, the $t$-dependence is \emph{polynomial} of degree 2: the series in Eq.~\eqref{eq:Veffallloops} must terminate at $n = 2$.

\section{Spin-1/2 fermions}
\begin{figure}
\includegraphics[scale=.8]{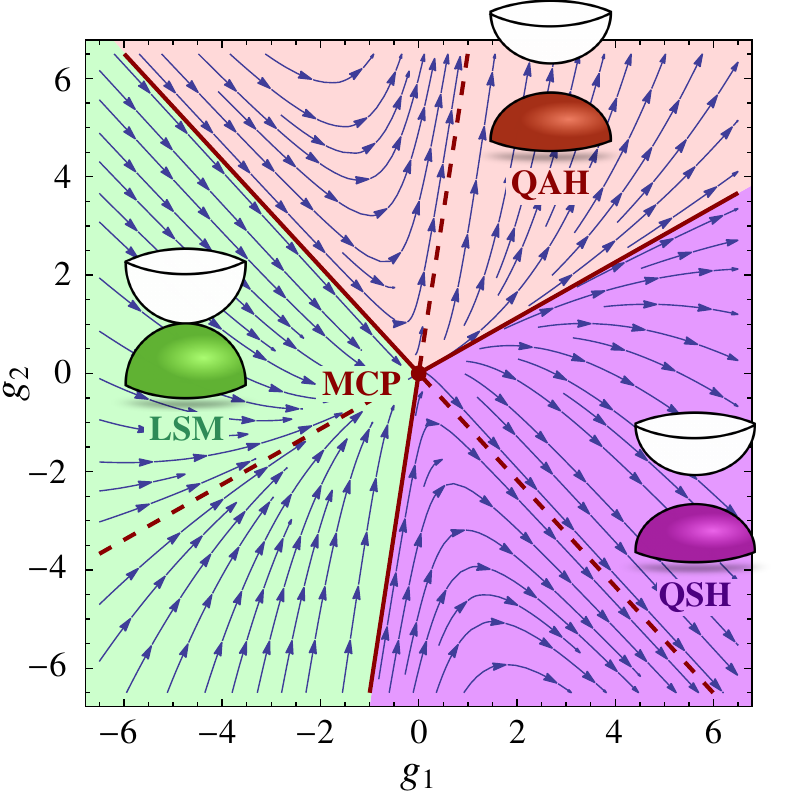}
\caption{Phase portrait in the $(g_1, g_2)$ coupling space assuming $g_0 = 0$, showing the Luttinger semimetal (LSM) and the two ordered states, viz.\ the quantum anomalous Hall (QAH) and the quantum spin Hall (QSH) states, and the multicritical point (MCP) at $(g_0, g_1, g_2) = (0, 0, 0)$.}\label{fig:phasediagspinhalf}
\end{figure}
For spin-$1/2$ fermions, we have four-component spinors, $\Psi = \left(\begin{smallmatrix}\psi_\uparrow\\ \psi_\downarrow\end{smallmatrix}\right)$, which is equivalent to $N=2$ flavors of two-component spinors. This leads to an increased number of independent four-fermion terms, among which the minimal set compatible with symmetries and closed under RG (at least at one loop) comprises \emph{two} terms:
\begin{align}
\mathcal{L}_{\text{int},\text{spin-1/2}} = -\frac{g_1}{2} \left(\Psi^\dagger (\sigma_a \otimes \mathds{1}_2) \Psi \right)^2 - \frac{g_2}{2} \left(\Psi^\dagger (\sigma_2 \otimes \mathds{1}_2)  \Psi \right)^2,
\end{align}
where $(\sigma_a) = (\sigma_3,\sigma_1)$ as before; the inverse propagator in $\mathcal{L}_0$ is trivially continued by taking the tensor product with $\mathds{1}_2$ for the spin degree of freedom: $\mathcal{G}_{0,N=2}^{-1} = \mathcal{G}_0^{-1} \otimes \mathds{1}_2$.
A third interaction term $g_0 (\psi^\dagger \psi)^2$ is allowed by symmetry, but will not get generated during the RG flow if absent initially~\cite{vafekspinhalf}; We postpone its discussion to the end of this section.
The RG flow of $g_1$ and $g_2$ is readily computed, and reads in $d = 2+\epsilon$ as
\begin{align}
 \frac{d g_1}{d\ln b} &= -\epsilon g_1 + \frac{1}{\pi}g_1^2 + \frac{1}{8\pi}g_2^2 - \frac{1}{4\pi}g_2 g_1, \label{eq:runningg1}\\
 \frac{d g_2}{d\ln b} &= -\epsilon g_2 + \frac{1}{2\pi}g_2^2 - \frac{1}{2\pi}g_1^2 + \frac{3}{2\pi} g_1 g_2. \label{eq:runningg2}
\end{align}
For $\epsilon = 0$, the above flow equations were derived previously in Ref.~\cite{vafekspinhalf}. In addition to the Gaussian fixed point $\text{G} \colon (g_1,g_2)_* = (0,0)$, there are interacting fixed points at
\begin{align}
\text{O}_1 \colon (g_1,g_2)_* &= (2.215\epsilon,-2.403\epsilon), \\
\text{O}_2 \colon (g_1,g_2)_* &= (0.6709\epsilon,4.374\epsilon), \\
\text{Q} \colon (g_1,g_2)_* &= (3.496\epsilon,1.977\epsilon).
\end{align}
Among these, $\text{G}$ is stable, $\text{O}_{i}$ ($i = 1,2$) have one stable direction and $\text{Q}$ is repulsive in all directions. For $\epsilon \to 0$, all three merge with $\text{G}$, rendering the latter multicritical. Importantly, the one-dimensional subspaces (i.e., lines) $\overleftrightarrow{\text{G}\text{O}_{i}}$ and $\overleftrightarrow{\text{G}\text{Q}}$ are RG invariant. %
This holds for all $\epsilon$, and in particular even at $\epsilon = 0$; it is precisely these lines which form the phase boundaries, shown in Fig.~\ref{fig:phasediagspinhalf} for the physical $d = 2$. %
The sector spanned by the rays $\overleftarrow{(-\text{O}_i)\text{G}}$ ($i = 1,2$) is the $N = 2$ incarnation of the Luttinger semimetal, in close analogy with the $N = 1$ case. Its complement in $(g_1,g_2)$ space is ordered; note, however, that there are in fact two ordered states corresponding to the fixed points $\text{O}_i$, separated by the ray $\overrightarrow{\text{G}\text{Q}}$. The ordered states can be classified by computing the fixed point ratios $(g_1/g_2)_*|_{\text{O}_i}$ and comparing with Ref.~\cite{vafekspinhalf}, which we have marked correspondingly in Fig.~\ref{fig:phasediagspinhalf}. The three different phases meet at the Gaussian fixed point. In fact, a simple calculation shows that the remaining third interaction channel $g_0 (\psi^\dagger \psi)^2$ is marginally relevant in the vicinity of this fixed point. In a generic microscopic model, the fixed point can thus in general be approached only by tuning two parameters (e.g., $g_0$ and $g_1$), corresponding to a multicritical point. However, at tree level all couplings are marginal, and the divergence of the correlation length hence still exhibits an essential singularity, while correlation functions at the quantum multicritical point display power-law decays with finite universal exponents, in close analogy with the spinless case.

